\documentclass[manuscript]{aastex631}

\usepackage{color}

\shorttitle{A candidate period of 4.605 day for FRB 20121102A and one possible implication of its origin}
\shortauthors{J. Li et al.}

\begin{document}


\title{A candidate period of 4.605 day for FRB 20121102A and one possible implication of its origin}


\author{Jixuan Li}
\affiliation{School of Physics and Astronomy, Sun Yat-Sen University, Zhuhai, 519082, Guangdong, China}
\author{Yang Gao}
\affiliation{School of Physics and Astronomy, Sun Yat-Sen University, Zhuhai, 519082, Guangdong, China}
\author{Di Li}
\affiliation{CAS Key Laboratory of FAST, National Astronomical Observatories, Chinese Academy of Sciences, Beijing 100101, China}
\affiliation{Research Center for Intelligent Computing Platforms, Zhejiang Laboratory, Hangzhou 311100, China}
\author[0000-0002-7568-8765]{Kinwah Wu}
\affiliation{Mullard Space Science Laboratory, University College London, Holmbury St.\;\!Mary, Surrey, RH5 6NT, UK}







\begin{abstract}
A firm establishment of the presence or the lack of periodicity in repeating Fast Radio Bursts (FRBs) 
is crucial for determining their origins.
Here we compile 1145 radio bursts of FRB 20121102A with fluence larger than 0.15 Jy ms
from observations using the Five-hundredmeter Aperture Spherical radio Telescope, Arecibo Observatory, Green Bank Telescope, Effelsberg Telescope, MeerKAT Telescope,
  Lovell Telescope, Deep Space Network 70 m radio telescopes, Very Large Array, and the Westerbork Synthesis Radio Telescope
  spanning the time interval of MJD 57175$-$58776.
A quasi-period of 157.1$_{-4.8}^{+5.2}$ day and a candidate quasi-period of 4.605$_{-0.010}^{+0.003}$ day are found through the phase-folding probability binomial analysis.
The former is consistent with previous findings and the latter is new.
The 4.605 day periodicity is more obvious in high-energy bursts with fluence larger than $1\times 10^{38}$ erg.
The presence of these (candidate) quasi-periods, 
  together with the corresponding width of burst accumulation in the phase space,
  are consistent with the bursts’ originating from a binary degenerate star system with a close-by planet around the primary neutron star.

\end{abstract}


\keywords{radio transit source --- radio bursts --- exoplanet: magnetic field --- time series analysis
}



\section{Introduction} 

High cadence  
  observations led to the discovery of fast radio bursts (FRBs),
  which usually have millisecond durations and are detected at frequencies between 110 MHz and 8 GHz
  \citep{Lorimer2007,Gajjar2018,Pleunis2021,Niu2022}.
More than 700 FRB sources have been identified, 
 with 50 repeaters among them \citep{CHIME2023}.
The host galaxies of some FRBs have been localized,
  the polarization and energy of bursts are measured,
  and the energy distribution and period of activities have been identified for a few repeating sources
  \citep{Petroff2022}.
FRBs are now believed to originate from a neutron star or magnetar
  through reconnection-driven free electron laser \citep{Lyutikov2020,Lyutikov2021},
  coherent curvature radiations \citep{Kumar2017,Lu2020,Wang2022},
  or synchrotron maser emissions \citep{Metzger2019}.
The finding of periodic activities from repeating sources suggests that
  the repeaters likely have progenitors in binary systems \citep{CHIME2020,Zhang2020C}
  or are originate from the precessions of flaring magnetars \citep{Levin2020}.

For the repeating source FRB 20121102A, a period of $\sim$160 day was found
  from extensive observations \citep{Rajwade2020,Cruces2021}.
The Five-hundredmeter Aperture Spherical radio Telescope (FAST)
  observations of this source revealed a bimodal burst energy distribution,
  hinting at different mechanisms or emission sites of the source \citep{Li2021}.
The time-domain behaviors may be linked to the energy distribution.
We thus carried out a comprehensive periodicity analysis of FRB 20121102A, based on data from
  the leading radio telescopes
  \citep{Spitler2016,Spitler2018,Law2017,Chatterjee2017,Scholz2017,Marcote2017,Hardy2017,Gajjar2018,Michilli2018,Hessels2019,Gourdji2019,Oostrum2020,Majid2020,Caleb2020,Pearlman2020,Rajwade2020,Cruces2021,Hilmarsson2021,Li2021,Hewitt2022,Jahns2023}.
 
The period of about 
 160 days detected in FRB 20121102A is a quasi-period with $>50\%$ duty cycle \citep{Rajwade2020,Cruces2021}.
This behavior is similar to radio bursts 
  from star-planet and Jupiter-satellite systems \citep[][]{Zarka2018}.
Searching for such quasi-periods  
  requires methods beyond the Fourier transformation in which period is the only parameter.
An improved methodology need to consider 
  the variable width of the bursts' distribution
  in the phase diagram.
The minimum string length method 
  developed in analyzing the period of variable stars allows such variations
  \citep{Dworetsky1983}.  
When studying the periodicity of bursts from exoplanet systems, 
  \citet{Gao2021} introduced the width of the burst duty cycle as another parameter in addition to the period.
They explored different candidate periods 
  by folding the time series of bursts to the phase space, 
  and then test different locations and widths of duty cycles within which the burst occurs.
By adopting this method,
  part of the bursts from 
  the star-planet system HD 189733 are found to be correlated with the planet.
The method is potentially applicable to repeating 
  FRBs, in which the burst duty cycle may also vary and affect the detection of periods in conventional periodograms.
In this paper, we further improve this method and apply it to FRB 20121102A.

The data are discussed in Section \ref{sec:data};
   in Section \ref{sec:ana} we present the quasi-period searching method and results of FRB 20121102A;
   implications of the burst origins are given in Section \ref{sec:discuss};
   conclusions are made in Section \ref{sec:conclu}.

\section{Data collection}
\label{sec:data}

There are two datasets of FRB 20121102A bursts used in this paper for the period analysis.
The first dataset (dataset I) summarizes bursts detected by 
  the FAST \citep{Li2021},
  Arecibo Observatory \citep[AO,][]{Spitler2016,Spitler2018,Chatterjee2017,Scholz2017,Marcote2017,Hardy2017,Michilli2018,Hessels2019,Gourdji2019,Hilmarsson2021,Hewitt2022,Jahns2023},
  Effelsberg Telescope \citep{Hardy2017,Spitler2018,Cruces2021,Hilmarsson2021},
  Very Large Array \citep{Law2017,Hilmarsson2021},
  Green Bank Telescope \citep[GBT,][]{Scholz2017,Gajjar2018,Michilli2018,Hessels2019},
  Lovell Telescope \citep{Rajwade2020},
  Westerbork Synthesis Radio Telescope \citep{Oostrum2020},
  MeerKAT telescope \citep{Caleb2020},
  and Deep Space Network telescopes \citep{Majid2020,Pearlman2020}.
In order to calculate the energy corrected burst rate in the period analysis (see Section \ref{subsec:method} for details),
  bursts with fluences larger than or equal to 0.15 Jy ms are kept in the list.
This means we have excluded bursts with energy smaller than this threshold,
  including the weak detections from 
  FAST \citep{Li2021},
  AO \citep{Spitler2016,Hilmarsson2021,Hewitt2022,Jahns2023},
  the Effelsberg Telescope \citep{Cruces2021,Hilmarsson2021},
  and the Green Bank Telescope \citep{Zhang2018}.
Additionally, the 12 Jy ms burst detected by Canadian Hydrogen Intensity Mapping Experiment (CHIME)
  telescope \citep{Josephy2019} is not included,
  because of its large fluence and short observation length that lead to an excessive count rate.
The first dataset finally includes 1145 bursts and can be found in the Supplementary data file,
  with an example shown in Table \ref{Table1}.
For each burst, the detection threshold of burst fluence, length of the observation session,
  and count rate corrected by the fluence threshold are given.

  \begin{table*}[h]
 \centering
\caption{Examples of FRB 20121102A bursts in dataset I.
   References: a. \citet{Spitler2016}, b. \citet{Scholz2016}, c. \citet{Li2021}. 
   A full version of this list can be found in the Supplementary data file.
 \label{Table1}}
  \begin{tabular}{@{}clccccll@{}}
  \hline\hline
  ID & MJD         & Ref. & Telescope & Frequency & Fluence $\geq$ & Observation &  Corrected    \\
     &             &      &           &   (MHz)   & (Jy ms)  & Length (hr)    &  Count Rate (ct/hr)  \\
  \hline
  1  & 57175.69314 & a    & Arecibo   & 1214-1537 & 0.15    & 1.75        &  0.57         \\
  2  & 57175.74351 & a    & Arecibo   & 1214-1537 & 0.15    & 1.75        &  0.57         \\
  3  & 57175.74566 & a    & Arecibo   & 1214-1537 & 0.15    & 1.75        &  0.57         \\
  4  & 57175.74762 & a    & Arecibo   & 1214-1537 & 0.15    & 1.75        &  0.57         \\
  5  & 57175.74828 & a    & Arecibo   & 1214-1537 & 0.15    & 1.75        &  0.57         \\
  6  & 57339.35604 & b    & GBT       & 1600-2400 & 0.15    & 0.83        &  1.20         \\
     &             &      &                &      &         &             &        \\

 1143& 58766.94890 & c    & FAST   & 1050-1450  & 0.15    & 1.00           &  1.00  \\
 1144& 58772.93022 & c    & FAST    & 1050-1450 & 0.15    & 1.00        &  1.00  \\
 1145& 58776.85096 & c    & FAST    & 1050-1450 & 0.15    & 1.00        &  1.00  \\
  \hline
  \hline
\end{tabular}
\end{table*}


The second dataset (dataset II) contains high-energy bursts with fluences $>5\times 10^{38}$ in FAST
 observations or equivalently $>0.4$ Jy ms in AO observations.
As a high-energy subset of dataset I, dataset II contains 289 bursts (cf. the Supplementary data file).
More details of the dataset can be found in \citet{Li2021}, \citet{Hewitt2022} and \citet{Jahns2023}.

\section{Quasi-periodicity analysis}
\label{sec:ana}

\subsection{Method}
\label{subsec:method}

Bursty phenomena like solar and stellar flares have higher occurrence rate in their `active regions',
  or when the binary/planet are in specific ranges of orbital positions \citep{Dulk1985,Perez2021}.
In such circumstance the bursts as a time series accumulate in relative phases of the solar/stellar rotation
  or binary/planet orbit.
Study of this kind of quasi-periodicity should target at two parameters, i.e.,
  the period and the width of the active sector in the phase space.
The method of finding these two parameters is essentially phase-folding,
  i.e., for one burst, its phase $P'$ when assuming a test period $T$ is calculated from the event MJD as 
  \begin{equation}
  {\rm MJD}-56000=CT+P'T,
  \label{equ:folding}
  \end{equation}
  where 56000 is subtracted from the MJD for convenience of computing, $C$ is an integer and $0\leq P'<1$.
We then introduce the test central active phase $P$ and its half width $\Delta P$ ($0<\Delta P\leq0.5$)
  and count the number of bursts with their phases
  \begin{eqnarray}
  P'\in[P-\Delta P,P+\Delta P) \qquad \qquad (P-\Delta P\geq0 \quad \& \quad P+\Delta P\leq1), \\
  \label{equ:phase}
  P'\in[0,P+\Delta P)\cup[P-\Delta P+1,1)\qquad \qquad\qquad \quad (P-\Delta P<0), \nonumber \\
  P'\in[0,P+\Delta P-1)\cup[P-\Delta P,1) \qquad\qquad\qquad \quad (P+\Delta P>1).
  \nonumber
  \end{eqnarray}
This number of bursts is noted as $N(T, P, \Delta P)$,
  which can be normalized by the total number of bursts, namely
  \begin{equation}
  \hat{N}(T, P, \Delta P)=\frac {N(T, P, \Delta P)} {N(T, P, 0.5)}.
  \label{equ:count}
  \end{equation}

For a phase-independent distribution of burst, $\hat{N}$ does not change when the test central active phase $P$ varies.
Contrarily, if $\hat{N}$ has variations above the random fluctuation and becomes significantly large for a test phase $P$
  in a certain test period $T$,
  the bursts likely have an intrinsic period $T$ and tend to occur around test phase $P$ within half sector width $\Delta P$.
Such burst accumulation can be quantitatively checked by assuming a phase-independent burst distribution
  and calculating the possibility that $\hat{N}(T, P, \Delta P)$ is equal to or larger than its current value.
This is simply the cumulative probability that $N(T, P, \Delta P)$ or more
  of the total $N(T, P, 0.5)$ bursts occur in the $2\Delta P$ phase range, i.e.,
  \begin{equation}
  F[N(T, P, \Delta P);N(T, P, 0.5), 2\Delta P]=\sum_{x=N(T, P, \Delta P)}^{N(T, P, 0.5)}B[x;N(T, P, 0.5), 2\Delta P],
  \label{equ:F}
  \end{equation}
  where
  \begin{equation}
  B(x;N,2\Delta P)=C_{N}^{x} (2\Delta P)^{x}(1- 2\Delta P)^{N-x}
  \end{equation}
   is the probability of occurrence of $x$ events in $N$ experiments each following the binomial distribution with single event probability $2\Delta P$
  (the whole phase range is unity).

In actual period searches, the cumulative probability $F$ defined in Equ.\ \ref{equ:F} can be calculated when one particular parameter set is tested.
As the number of independent periods (frequencies), central active phases and phase widths 
  searched increases, the minimum value of $F$ becomes smaller compared to in one particular test.
We take into account this effect by following the definition of false alarm probability ($FAP$) in
  \citet{Horne1986} and \citet{VanderPlas2018}, i.e.,
  \begin{equation}
  FAP=1-(1-F)^{N_{\rm P}N_{\rm \Delta P}N_{\rm eff}}.
  \label{equ:FAP}
  \end{equation}  
Here $F$ is calculated using Equ.\ 4 for a particular test, $FAP$ is the probability that the detected burst accumulation in the test would be achieved under the phase-independent burst assumption.  $N_{\rm P}$ and $N_{\rm \Delta P}$ are the numbers of $P$ and $\Delta P$ searched in the trials, close to the independent values of these two parameters; $N_{\rm eff}$ is the number of independent frequencies which can be calculated from
  \begin{equation}
  N_{\rm eff}=f_{\rm Ny}\Delta T,
  \end{equation}
  with $f_{\rm Ny}$ being the Nyquist frequency and $\Delta T$  the maximum time interval in the period search. 
For irregularly spaced time series, the Nyquist frequency 
  can be estimated as \citep{Koen2006}
  \begin{equation}
  f_{\rm Ny}=\frac{0.5}{\delta _*},
  \end{equation}
with $\delta _*$ being the precision of time measurements.
After calculating $FAP$, the confidence (sigma) level $Z$ can be estimated through the distribution function of the standard Gaussian distribution \citep{Carpano2007,Forbes2011}, i.e.,
  \begin{equation}
  1-FAP\approx \frac{1}{\sqrt{2 \pi}} \int_{-Z}^{Z} e^{-\frac{x^{2}}{2} } dx.
  \end{equation}
If $FAP$ is in 
  close proximity to zero, 
  the phase-independent assumption of bursts can be ruled out
  and the intrinsic accumulation of bursts with period $T$, central active phase $P$ and half sector width $\Delta P$ can be confirmed at confidence level $Z$.
The above method
  is named Phase-folding probability Binomial Analysis (PBA).

Before adopting the above periodicity analysis, we make two additional corrections on observation length and detection threshold when counting the bursts of FRB 20121102A.
Firstly, the uneven and sometimes periodic distribution of observation window and its width may lead to pseudo periods of the bursts.
However, the effect of the width of the window can be partially avoided by using the count rate
  instead of the number of bursts during a continuous observation session.
The count rate (around central active phase $P$ with half sector width $\Delta P$) for bursts observed during
  an continuous session i with length of observation time $\tau_{\rm i}$ is
  \begin{equation}
  n_{\rm i}(T, P, \Delta P)=\frac {N_{\rm i}(T, P, \Delta P)}{\tau_{\rm i}},
  \end{equation}
  where $N_{\rm i}(T, P, \Delta P)$ is the number of bursts detected in this observation session.
The total count rate in an observation campaign  is the sum of the above rate over all observation sessions, i.e.,
  \begin{equation}
  n(T, P, \Delta P)=\sum_{\rm i}n_{\rm i}(T, P, \Delta P).
  \end{equation}
The normalized count rate is accordingly
  \begin{equation}
  \hat{n}(T, P, \Delta P)=\frac {n(T, P, \Delta P)} {n(T, P, 0.5)}.
  \end{equation}

Secondly, when considering bursts detecteded by different telescopes, they usually have different lower limits of fluences.
In the same observation session toward a repeating source, bursts recorded by a telescope with higher fluence threshold
  will be less than by a telescope with lower fluence threshold.
This bias can be corrected for FRB 20121102A as we know the energy distribution of its bursts.
All bursts in dataset I have fluences equal to or higher than 0.15 Jy ms,   above which the generalized Cauchy
  function of energy distribution in \cite{Li2021} can be simplified to a constant + a power law.
For bursts with fluences between 0.15 and 0.24 Jy ms, corresponding to $\sim 2$ and 3 $\times 10^{38}$ erg in FAST observations,
  the burst rate can be treated as a constant \citep[Fig.\ 2 in][]{Li2021}.
For bursts with fluences larger than 0.24 Jy ms, the energy distribution follows a power law
  \citep[Table 1 of][]{Li2021}.
So if we assume this distribution also applies for bursts detected by other telescopes, the energy distribution function for dataset I is \footnote{
The distribution function is achieved assuming the detection rates are normalized by the rate at fluence of 0.15 Jy ms. It is also noted that the FAST survey is complete for fluences larger than 0.15 Jy ms \citep{Li2021}.}
  \footnote{
The variation of energy 
  distribution with burst frequency is not considered here, which may lead to over or less estimate of the normalized equivalent count rate for the bursts detected in frequencies other than 1.05 GHz to 1.45 GHz.}
  \begin{eqnarray}
  p(E) & =p(0.24) \qquad \qquad (0.15 \leq E <0.24 ), \\
  \label{equ:phase}
  p(E) & =p(0.24)\left(\frac{E}{0.24}\right)^{-1.37} \qquad ( E \geq 0.24 ), \nonumber
  \end{eqnarray}  
where $E$ is the burst energy in Jy ms.
To correct the count rate of bursts detected by telescopes at different fluence thresholds,
  we further integrate the above distribution function from the threshold energy $E$ to $+\infty$ and get 
  \begin{eqnarray}
  f(E) & =\int_E^{+\infty}p(E')~{\rm d}E'=\frac{0.89-E}{0.74} \qquad \qquad (0.15 \leq E <0.24 ), \\
  \label{equ:phase}
  f(E) & =\int_E^{+\infty}p(E')~{\rm d}E'=\frac{0.24^{1.37}}{0.74\cdot0.37\cdot E^{0.37}} \quad\qquad ( E \geq 0.24 ), \nonumber
  \end{eqnarray} 
  which is the detection rate (normalized by $f(0.15)$) for bursts with energy higher than $E$.
We adopt it and calculate from the count rate of an observation j with threshold energy $E_{\rm j}$ to
  get the equivalent count rate with threshold $0.15$ Jy ms, i.e.,
  \begin{equation}
  n_{0.15,E_{\rm j}}(T, P, \Delta P)=\frac {n_{E_{\rm j}}(T, P, \Delta P)} {f(E_{\rm j})}.
  \end{equation}
Summing the above rate over all observations with their corresponding fluence thresholds,
  \begin{equation}
  n_{0.15}(T, P, \Delta P)=\sum_{\rm j}n_{0.15,E_{\rm j}}(T, P, \Delta P)
  \end{equation}
  is the total count rate for a dataset, equivalently at $0.15$ Jy ms. Its normalized form is
  \begin{equation}
  \hat{n}_{0.15}(T, P, \Delta P)=\frac {n_{0.15}(T, P, \Delta P)} {n_{0.15}(T, P, 0.5)}.
  \end{equation}

As a summary, practically each burst is counted as one divided by its observation length $\tau_{\rm i}$
  and the energy correction function $f(E_{\rm j})$ in the PBA procedure.
For dataset I, we do the period analysis based on $\hat{n}_{0.15}(T, P, \Delta P)$,
  the normalized equivalent count rate with threshold $0.15$ Jy ms.
For dataset II, since all detections have the same energy 
    threshold, the period analysis is carried out by simply accounting the observation window function to the probability for a single burst to occur within a phase range.
That is, by replacing $2\Delta P$ in Equ.\ \ref{equ:F} with the normalized 
  observation window function $W(T, P, \Delta P)$, i.e., the normalized observation time spent within the region with central active phase $P$ and half width $\Delta P$, the PBA is made upon the number of bursts $N_{\rm i}(T, P, \Delta P)$.

\subsection{Result}

\subsubsection{Period of 157.1 day}

\begin{figure}
 \epsscale{1.0} \plotone{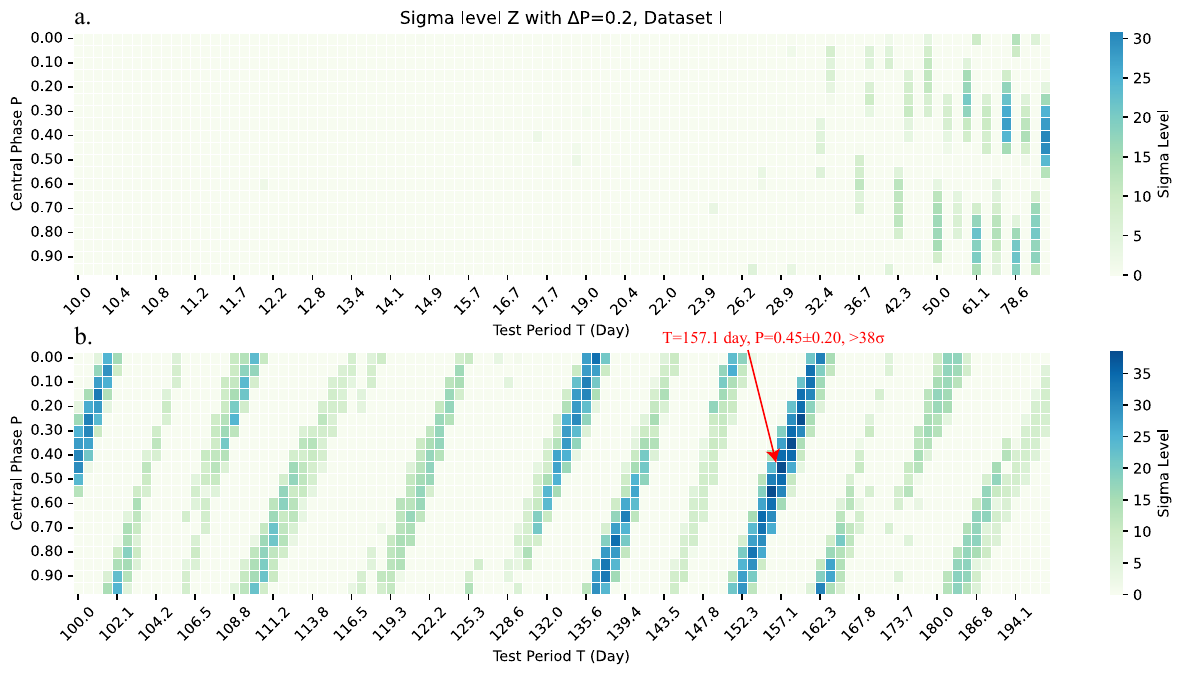}
\caption{The distribution of sigma level $Z$ for test periods from $10$ to $100$ day (\textbf{a}) and $100$ to $200$ day (\textbf{b}), with test central active phase from $0$ to $0.95$ and half width $\Delta P=0.2$, calculated for dataset I. The color of pixels show the sigma level $Z$. High sigma level is found in $T=157.1$ day at central active phase $P=0.45$ with $Z>38$. The presence of the high sigma level pixels indicates the existence of a period around 157.1 day.}
\label{fig:HM1}
\end{figure}

\begin{figure}
 \epsscale{0.8} \plotone{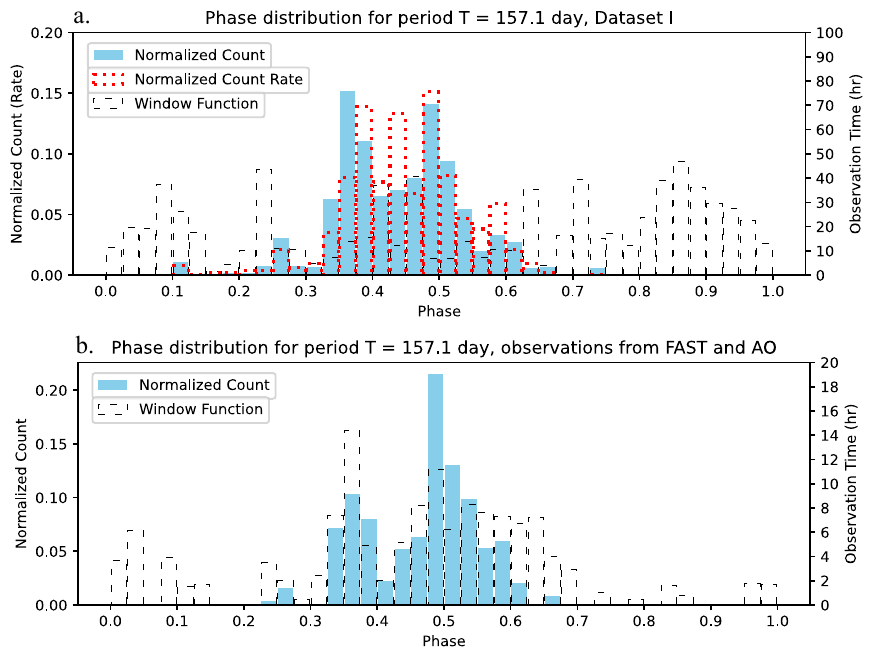}
\caption{For period 157.1 day, the phase distribution is plotted for dataset I (\textbf{a}) and observations from FAST and AO (\textbf{b}). The blue, dotted red and dashed black histograms are the normalized burst counts $\hat{N}$ (cf. Equ. 3), normalized count rate $\hat{n}_{0.15}$ (cf. Equ. 17) and observation window respectively. Almost all the bursts concentrate at phases 0.325 to 0.625, and most of the observations from FAST and AO are made within this duty cycle (\textbf{b}). }
\label{fig:PD157}
\end{figure}

The $FAP$ 
 is first calculated based on dataset I.
Test frequencies from $10^{-1}$ to $100^{-1}$ day$^{-1}$ and $100^{-1}$ to $200^{-1}$ day$^{-1}$ with equal frequency steps of  $1000^{-1}$ day$^{-1}$ and $20000^{-1}$ day$^{-1}$ are explored respectively.
For each test frequency, test central active phases from $P$=0.00 to 0.95 with 0.05 step are adopted,
covering the entire phase space if the width of the active sector is not smaller than 0.05.
The distribution of confidence level $Z$ for half active phase $\Delta P$=0.2 is shown
  in Fig.\ \ref{fig:HM1}.
It is noted that high-sigma confidence levels occur around period $T$=157.1 day and central active phase $P$=0.45.
For the test period $T=157.1$ day,
the high-sigma level squares spread to segments covering $20 \%$ of the entire phases,
  demonstrating the accumulation of bursts being in a wide phase range. 
The effect of observation window function is excluded from this period \citep[Fig.\ 2, see also][]{Cruces2021}.
An uncertainty of $_{-4.8}^{+5.2}$ days can be estimated,
out of which the sum of normalized count rate within the active phase decreases to be $\hat{n}_{0.15}(T, P, 0.2) < 0.90$, significantly smaller than the value for 157.1 day in the $\Delta P$=0.2 trial.
The phase distributions of the normalized count (rate) and window function for the 157.1 day period are shown
  in Fig.\ \ref{fig:PD157}.
The window function is the observation length at each continuous observation,
  with data extracted from \citet{Spitler2016,Spitler2018,Scholz2016,Scholz2017,Law2017,Michilli2018,Hardy2017,Gajjar2018,Cruces2021,Majid2020,Caleb2020,Pearlman2020,Rajwade2020,Li2021,Hewitt2022,Jahns2023}.
It is seen that the count peaks at phases between 0.35 and 0.6, with bursts spread to $\sim$50$\%$ phases in dataset I.
Most of the observation windows of FAST and AO are within the duty cycle (cf. panel b).

However, high-sigma level squares are found in other periods/phases as well according to Fig.\ \ref{fig:HM1}.
Do they also indicate intrinsic periods or are they caused by the 157.1 day period?
From Equ.\ \ref{equ:folding} when we slightly change the test period from $T$ to $T+\delta T$, the change of the phase for a burst is
  \begin{equation}
  \delta P'\approx -\frac{C}{T}\delta T.
  \label{equ:phase}
  \end{equation}
According to Table \ref{Table1}, for the 157.1 day period, cycle $C$ is between 7 and 17 for bursts in dataset I.
Bursts in each cycle have a specific `curve' in Fig.\ \ref{fig:HM1},
  and they meet at the high-sigma squares around $T=157.1$ day,
  indicating the existence of an intrinsic period for bursts in different cycles.
They gradually disperse in the phase space when the test period goes away from the intrinsic period of 157.1 day.
So the phase concentrations for $T>\sim$ 80 day in Fig.\ \ref{fig:HM1} (mainly in panel b) are likely due to the intrinsic period of 157.1 day.
Also, trial periods $>\sim$10 days (panel a) have been studied at a high confidence level in \citet{Rajwade2020} and \citet{Cruces2021}
  where no other intrinsic period is found.
We then seek for possible periods between 2 and 
10 days in this paper.

\subsubsection{Period of 4.605 day}

\begin{figure}
 \epsscale{1.0} \plotone{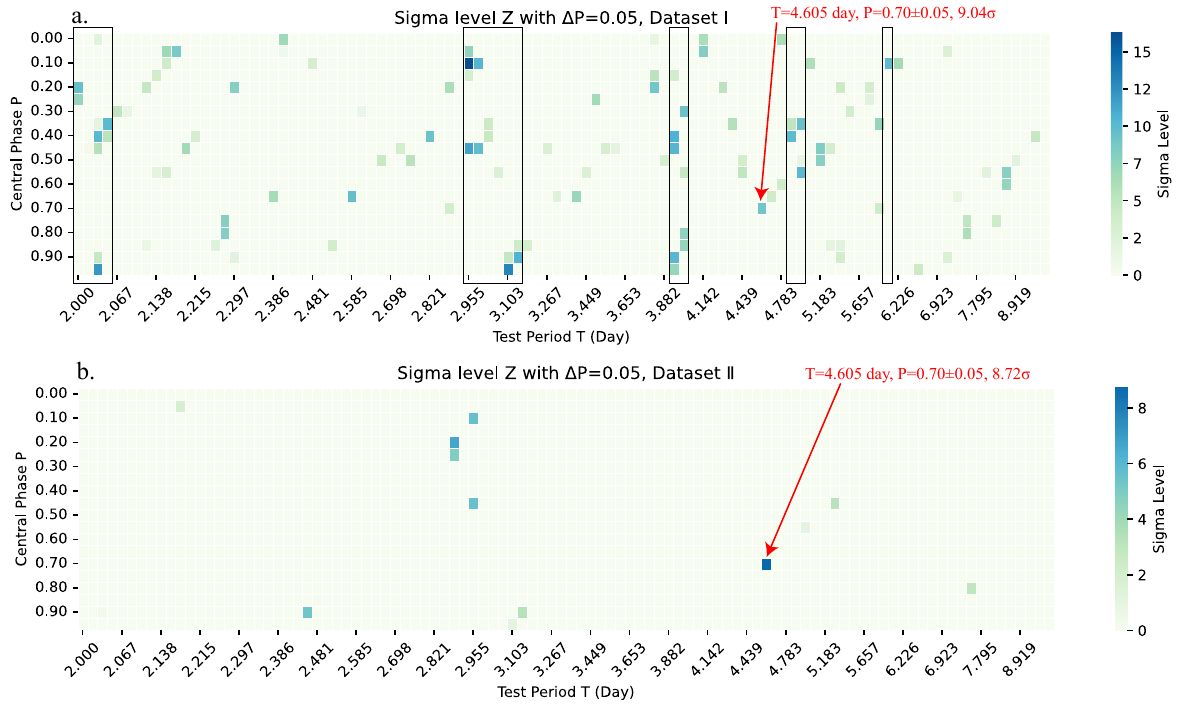}
\caption{The distribution of sigma level $Z$ for test periods from 2 to 10 days and test central active phase from 0 to 0.95, calculated for dataset I (\textbf{a}) and dataset II (\textbf{b}). The color of pixels show the sigma level  $Z$. The black-edge boxes in panel \textbf{a} show the results where $T$ is close to integers of 2, 3, 4, 5 and 6 days. For $T=4.605$ day, high sigma level $Z$ = 9.04 and 8.72 are found at $P$ = 0.70 for dataset I and dataset II respectively.}
\label{fig:HM2}
\end{figure}

\begin{figure}
 \epsscale{0.8} \plotone{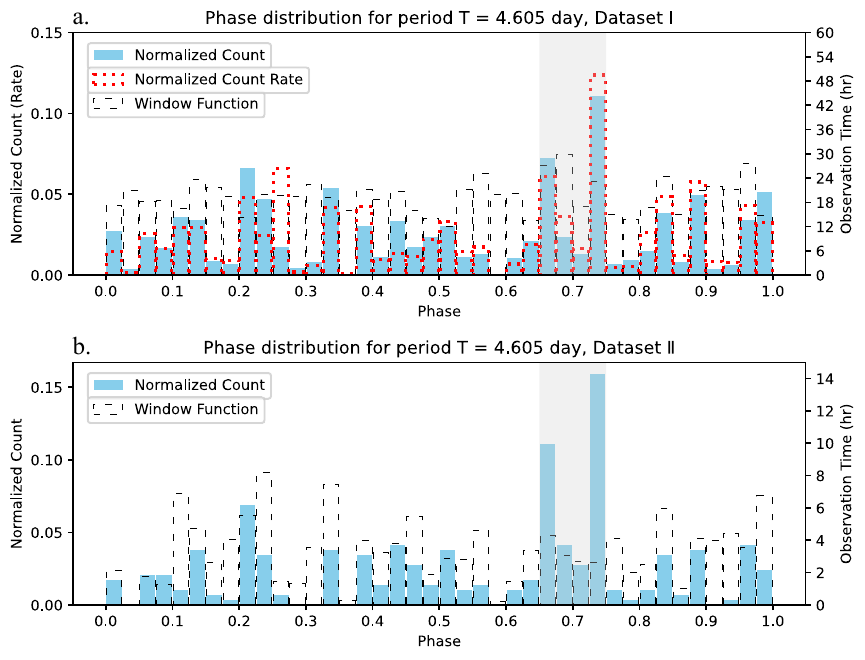}
\caption{For period 4.605 day, the phase distribution is plotted for dataset I (\textbf{a}) and II (\textbf{b}). The blue, dotted red and dashed black histograms are the normalized burst counts $\hat{N}$ (cf. Equ. 5), normalized count rate $\hat{n}_{0.15}$ (cf. Equ. 17) and observation window respectively. The active phase is shown by the gray region. Part of the bursts concentrate at phase $0.70\pm0.05$ for both datasets.}
\label{fig:PD4.6}
\end{figure}

The PBA method is carried out for test frequencies from $2^{-1}$ to $10^{-1}$ day$^{-1}$ with an equal frequency step of $250^{-1}$ day$^{-1}$
(corresponding to test period from $T$=2 to 10 day with approximately $\sim$0.01 day step).
For each test period, test central active phases from $P$=0 to 0.95 with 0.05 step are adopted,
  and half active phase $\Delta P$=0.025 to 0.2 with step size 0.025 is considered.
The results with half width $\Delta P$=0.05 for datasets I and II are shown in Fig.\ \ref{fig:HM2}.
For dataset I, there are clearly 2, 3, 4, 5 and 6-day periods arising from the approximately one-day period of observation window (Fig.\ \ref{fig:HM2} \textbf{a}).
A local high-sigma square with test period $4.605$ day is identified in dataset II (Fig.\ \ref{fig:HM2} \textbf{b}).
At this test period, a high confidence level pixel is also found for dataset I
(Fig.\ \ref{fig:HM2} \textbf{a}) 
  with the same central active phase,
  suggesting the existence of an intrinsic period of $4.605$ day.
For the totally 1145 bursts in dataset I, 252 bursts occur in the $0.70\pm0.05$ phase range, for which $FAP=7.97\times 10^{-20}$, corresponding to a 9.04$\sigma$ confidence level.
For dataset II, there are 98 of the totally 289 bursts detected within the $0.70\pm0.05$ phase range, 
with the false alarm probability $FAP=1.35\times 10^{-18}$, corresponding to 8.72$\sigma$.
The uncertainty of the period is estimated to be $_{-0.010}^{+0.003}$ day,
  out of which the confidence level decreases to less than $4 \sigma$ for both datasets I and II.

\begin{figure}
 \epsscale{1.1} \plotone{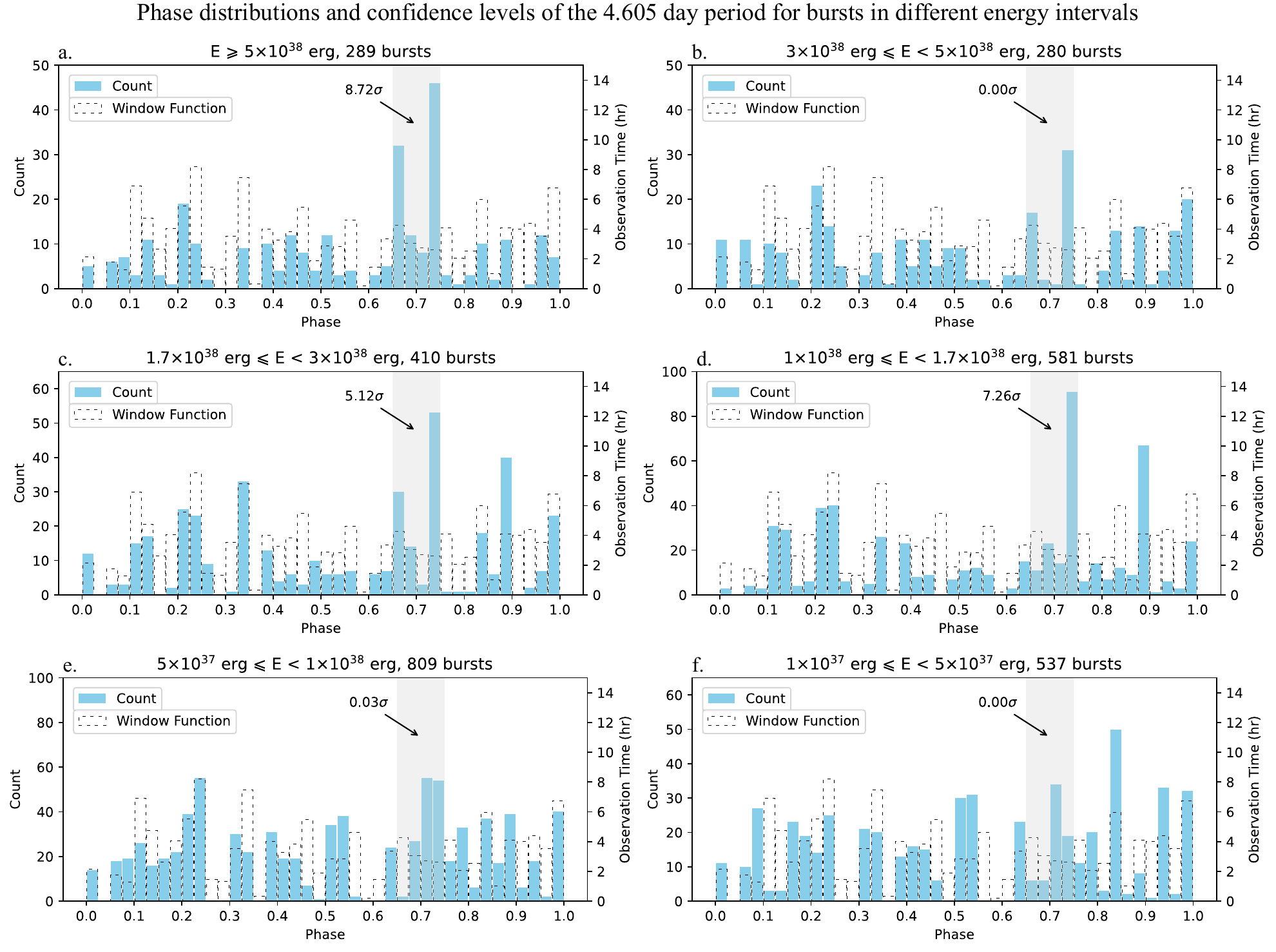}
\caption{Phase distribution for the 4.605-day period for bursts observed with FAST and AO in different fluences. The blue and dashed black histograms show the burst counts and observation windows respectively.
The active phase is shown by the gray region, with the confidence level of the 4.605-day period indicated.
}
\label{fig:ET}
\end{figure}

The phase distribution for period $4.605$ day is shown in Fig.\ \ref{fig:PD4.6}.
For both datasets, phase distributions show peaks at the active phase $0.70\pm0.05$, with the concentration of bursts being more obvious for dataset II containing high energy bursts (Fig.\ \ref{fig:PD4.6} \textbf{b}).
To find the lower energy cut-off of the 4.605-day period, 
we divide the bursts observed by FAST and AO into six groups according to their fluences.
The confidence level of the 4.605-day period is calculated for each group of bursts, shown in Fig.\ \ref{fig:ET} with corresponding phase distributions.
High confidence levels are reached when the burst fluence is higher than $5\times 10^{38}$ erg (Fig.\ \ref{fig:ET} \textbf{a}) or between $1\times 10^{38}$ erg and $3\times 10^{38}$ erg (Fig.\ \ref{fig:ET} \textbf{c} and \textbf{d}).
There is a drop of confidence level for bursts with fluences between $3\times 10^{38}$ erg and $5\times 10^{38}$ erg, but the concentration of bursts can still be seen at the same phase interval (Fig.\ \ref{fig:ET} \textbf{b}).
For bursts with fluences lower than $1\times 10^{38}$ erg, 
the confidence level decreases to $\sim$0 and the concentration of bursts is no longer seen (Fig.\ \ref{fig:ET} \textbf{e} and \textbf{f}).
Therefore, the 4.605$_{-0.010}^{+0.003}$ day period is an intrinsic property for bursts
with fluences higher than $1\times 10^{38}$ erg.
This low energy cut-off of periodicity is roughly consistent with the $3\times 10^{38}$ erg threshold of bimodal energy distribution in \cite{Li2021}.

Furthermore, this period is a quasi-period in the sense that only part of the bursts concentrate in the active phase. This is similar to the radio bursts from the interactions of Jupiter with its satellites Io and Ganymede, which have periods related to the orbits of the satellites. The orbital periods only applies to  $\sim 30\%$ (bursts related to Io) and $\sim 6\%$ (bursts related to Ganymede) of all observed bursts from the Jovian system, with the rest from Jupiter alone \citep{Zarka2018}.  
To summarize, the accumulation of bursts shown in Fig.\ \ref{fig:PD4.6} is a rare event if the bursts distribute evenly with phase.
So we conclude at a confidence level higher than 8$\sigma$ the existence of the candidate period 4.605 day and the accumulation of part of the bursts in the specific active phase range of $0.70\pm0.05$.
The reason it is called a `candidate' period is that the sampling effect of discrete observations should be checked by future observations.

\begin{figure}
 \epsscale{1.0} \plotone{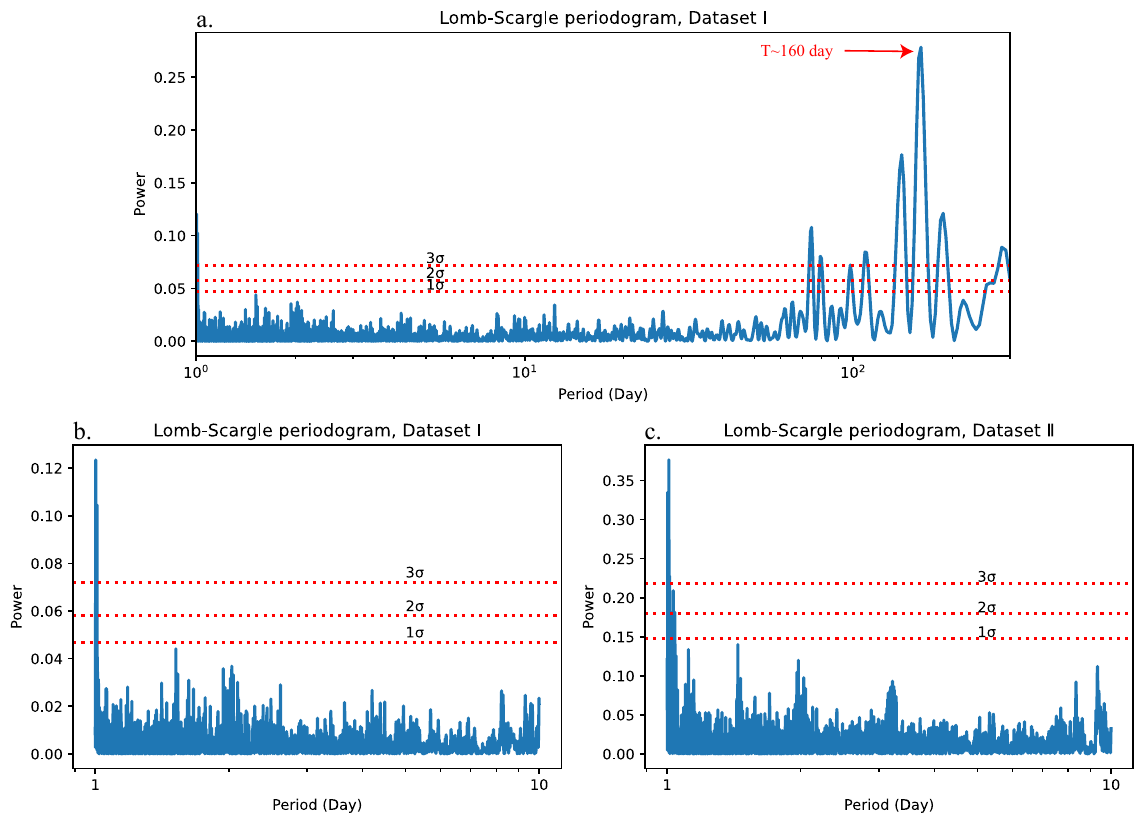}
\caption{Lomb-Scargle periodogram of dataset I with test periods from 1 to 300 days (panel \textbf{a}),
  and of dataset I (panel \textbf{b}) and dataset II (panel \textbf{c}) with test periods from 1 to 10 days.
The periodogram of dataset I (panel \textbf{a}) shows a peak near period $T$ = 160 day,
  in agreement with the result of \citet{Rajwade2020} and \citet{Cruces2021}.
  In both panel \textbf{b} and \textbf{c} we can find the one-day period arising from the observation window but no peak around period $T$ = 4.6 day is found.
}
\label{fig:LS}
\end{figure}

\begin{figure}
 \epsscale{1.1} \plotone{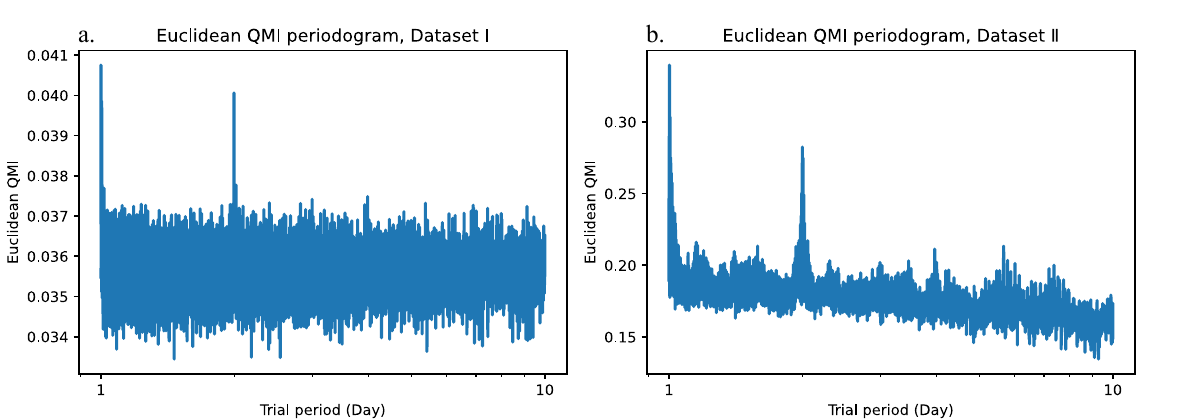}
\caption{Euclidean QMI estimators of dataset I (panel \textbf{a}) and dataset II (panel \textbf{b}) with test periods from 1 to 10 days. 
We can see 1- and 2-day periods of the observation window in both panels, but no peak around period $T$ = 4.6 day can be found.
}
\label{fig:QMI}
\end{figure}

We have also explored the periodicity of bursts near $T$=4.6 days using other methods.
The $\chi$-square test \citep{Leahy1983} and H test \citep{deJager1989,deJager2010}
give $\chi$-squares and $H$s larger than the average near 4.6 day,
but all of these values are much lower compared to those in the vicinity of integer numbers of days.
These results are consistent with the PBA result of 4.605 day, but cannot be considered as candidate periods from the methods themselves due to the existence of many other local peaks. 
The Lomb-Scargle (LS) periodogram \citep{VanderPlas2018} for dataset I detects a period at 160 day (Fig. \ref{fig:LS} \textbf{a}), which is consistent with the result of \citet{Rajwade2020} and \citet{Cruces2021}.
However, no period can be found between 1 to 10 day using LS, except the one-day period from observation windows (Fig. \ref{fig:LS} \textbf{b} and \textbf{c}).
The quadratic mutual information (QMI) analysis \citep{Huijse2018}, generated by python package frbpa\footnote{Available at github.com/KshitijAggarwal/frbpa .}, shows one-day and two-day periods of the observation window, with no period near 4.6 day found (Fig. \ref{fig:QMI} \textbf{a} and \textbf{b}).
Wavelet analysis have also been tested for datasets I and II, but no period is found.
The period of 4.605 day has a small duty cycle of 0.10 as seen from Fig.\ \ref{fig:PD4.6}, making it hard to be detected in the wavelet analysis, Lomb-Scargle periodogram and QMI analysis, which are based on the Fourier transform.
Probability analysis methods such as the $\chi$-square test, H test and PBA could be more applicable to periods of bursty phenomena with small duty cycles.
Particularly, the PBA method is more sensitive to such quasi-periods as it is in principle a  $\chi$-square test, but more specifically based on the single active-phase assumption.

\begin{figure}
 \epsscale{1.1} \plotone{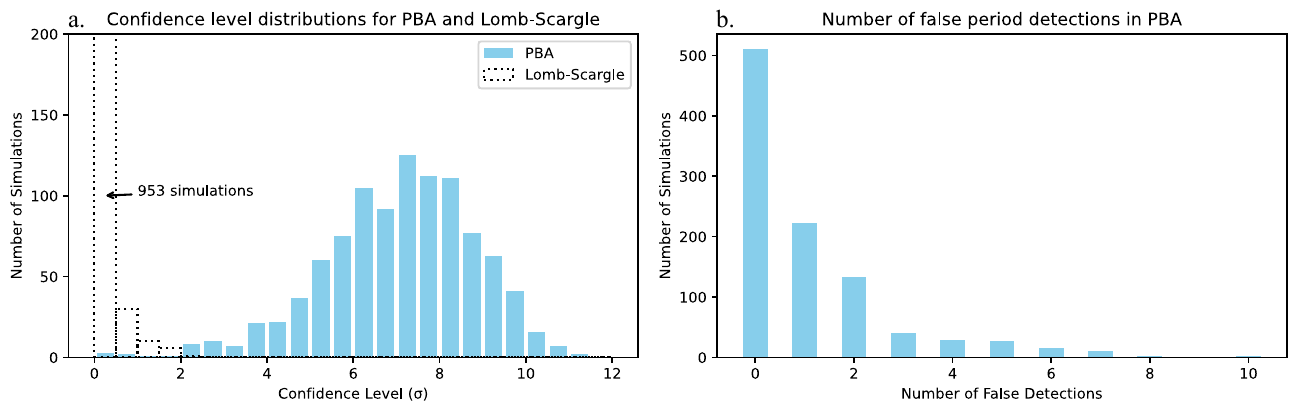}
\caption{Results of PBA and Lomb-Scargle (LS) periodograms on burst data generated by Monte Carlo simulation. One thousand sets of burst data are generated by assuming $28\%$ bursts have 0.1 duty cycle, 4.605-day periodicity and $72\%$ are random and have no periodicity. In panel \textbf{a}, the blue and dotted black histograms are the confidence levels for the 4.605-day period in PBA and LS, respectively. Panel \textbf{b} shows the number of false period detections in PBA, with confidence levels higher than that of the 4.605-day period. The PBA method is more sensitive in detecting the 4.605-day period without much confusion from false periods.
}
\label{fig:MC}
\end{figure}

To quantify the different sensitivities of PBA and LS periodograms, we generate a thousand sets of burst data through Monte Carlo simulation and apply the two methods to the data sets. 
Taking dataset II as a reference, each set of simulation data contains $28\%$ bursts with 0.1 duty cycle, 4.605-day periodicity, and $72\%$ random bursts with no periodicity.
The same observation window of dataset II is then applied to the simulation data, 
  resulting in `detected' bursts ranging from 250 to 310 for each data set.
The confidence levels of the 4.605-day period are calculated using PBA and LS respectively,
  with the distributions shown in Fig.\ \ref{fig:MC} \textbf{a}.
The LS periodogram always gets a $<1\sigma$ confidence level,
while PBA detects the 4.605-day period at confidence levels between 4 and 10.
We also show in Fig.\ \ref{fig:MC} \textbf{b} the number of false detections of periodicity in PBA. 
A false detection is the detection of a period other than 4.605 day, at which the confidence level is higher than that of 4.605-day.
In $>\sim85 \%$ sets of simulation data, there are less than 3 false detections.
So the detection of the 4.605-day period by PBA is seldom confused by false detections.
The above results show that PBA is more sensitive than LS when part (28$\%$ here) of the bursts have small duty cycle (0.1 here) periodicity.


\section{Discussions on the burst origin and formation of FRB 20121102A} 
\label{sec:discuss}

\subsection{Possible origin of the radio emission} 


Young magnetars are active.  
Their rapid rotation and strong magnetic field 
  power wind outflows, 
  and blast waves can also be launched 
  from them into the wind outflows. 
When the blast waves interact with the out-flowing materials, 
  shocks will be formed. 
The shocks accelerate particles, 
  leading to the radio emission as seen in the bursts from these systems. 
This ``shocks in magnetar outflows'' is the scenario that 
  \cite{Beloborodov2017,Beloborodov2020} 
  had put forward in explaining the origin of FRBs.
For FRB 20121102A, the magnetic field is required to be 
 $\geq 10^{14}$ Gauss  \citep{Cheng2020}.    
With periodicities between a dozen and more than one hundred days 
  found in repeating FRBs, 
  there is now a consensus that the burst periodicity 
  is associated with either the orbital periods of binary systems 
  with the companion being a degenerate star or massive star, or the precession of a magnetar \citep{Levin2020,Du2021,rajwade2023}.
And the possible origin of
  a ultra-long period 
  neutron star cannot be ruled out for FRBs with shorter periods \citep{kramer2023}.  

For FRB 20121102A, a highly eccentric binary orbit with a critical 
  separation r$_{\rm c}\sim7\times 10^{13}$ cm
  (or 4.67 AU)
  between the binary degenerates is assumed to account for its period of $\sim160$ days and a 50$\%$ burst duty cycle \citep{Du2021}. 
We adopt this geometric picture in which magnetised plasma results from the accretion flow from the companion white dwarf,
  spreads between the companion and the neutron star when they have separations smaller than r$_{\rm c}$.
One possible explanation of the 4.605 day period is the existence of a planet close to the neutron star.
For a neutron star of $1.4~{\rm M}_{\odot}$, 
  this period therefore implies a planet orbital separation of 
  $0.061$ AU.  
Also, the planet will reside in the neutron star's magnetosphere 
  and interact with the wind from the neutron star 
  and the inflow material when the neutron star accretes.   
As such, the interaction between the planet 
  and the magnetised plasma around it would form Alfv\'{e}n wing instability,
  which gives rise to radio pulses, 
  contributing to the quasi-periodic variations 
  of single pulses
  observed in the FRB \citep{Mottez2014,Mottez2020}.  
Relativistic beaming makes the high intensity radio bursts focus at a narrow range of direction.
When the radiation direction passes through the line of sight in each planet orbit,
  an increasing number of radio bursts are expected to be seen by the observer.
Recalling the binary degenerate star system with the critical separation r$_{\rm c}$,
  only when the binary interaction is on can there be accretion flows 
around the neutron star 
  that hosts the planet, 
  such that the planet 
  is able to interact with the plasma around it to produce 
  burst like radio emission.   
In this scenario, 
  using a model with two binary degenerate star, 
  of which the neutron star has a closely orbiting planet, 
  we are able to explain the $\sim160$ day period with 50$\%$ duty cycle,
  and the the 4.605 day period 
  for part of the high energy bursts 
  in a consistent manner. 
  
 \begin{figure}
 \epsscale{0.6} \plotone{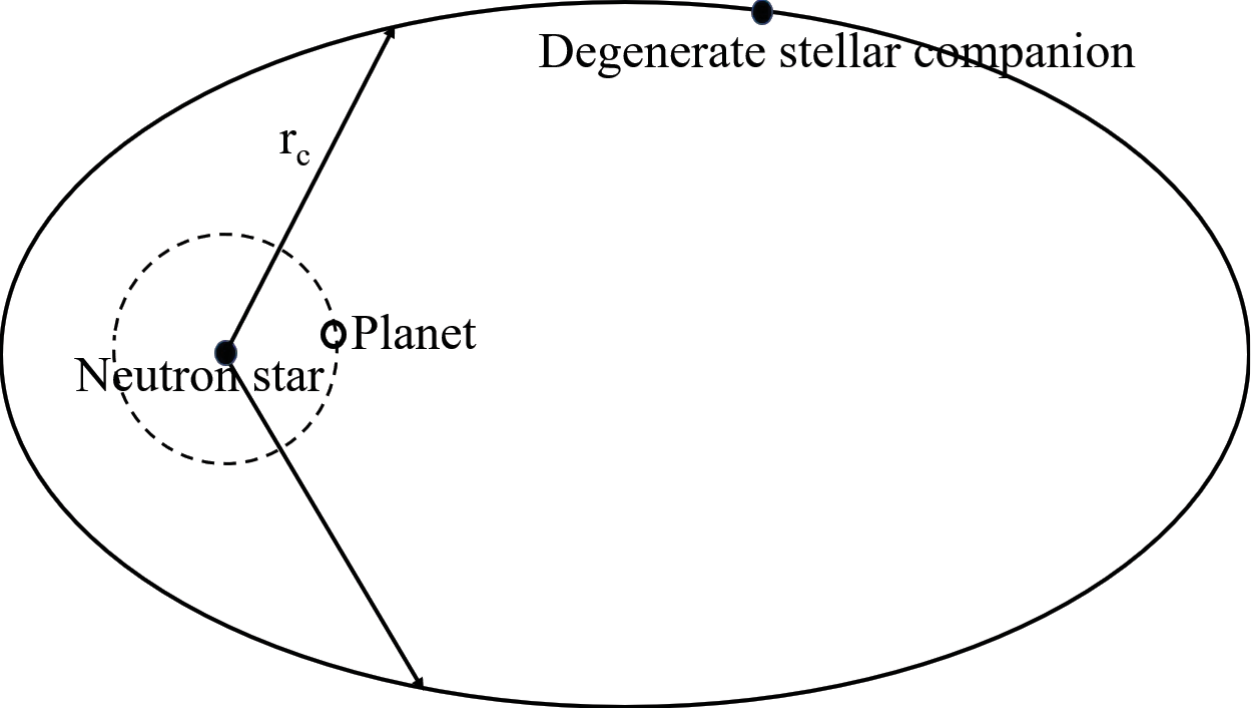}
\caption{Schematic of possible origin of FRB 20121102A: a binary degenerate star system with a close-by planet around the primary neutron star. The binary has a highly eccentric orbit, with r$_{\rm c}$ the critical separation between the primary neutron star and the degenerate stellar companion under which there are stellar wind or accretion interactions between the binary \citep{Du2021}. The close-by planet around the primary neutron star has an orbit period of $\sim$4.6 day. When there is accretion flow it fills the space between the neutron star and the planet. In each planet orbit, when the radiation induced by the planet passes through the line of sight, high intensity radio bursts focusing at a narrow range of direction are detected \citep{Mottez2020}.}
\label{fig:origin}
\end{figure}
  
Apart from the above scenario, 
  there are other possible mechanisms for the bursts
  under the same geometric setting shown in Fig.\ \ref{fig:origin}.  
We discuss them as supplementary models.
The first alternative is magnetic reconnection.
Magnetic field reconnections, 
  as those observed in the solar system space environment 
  \citep[see e.g., ][for a review]{Hesse2020},  
  convert the energy stored in the magnetic fields 
  to the kinetic energy of particles.  
Such high energy particles 
  can form Alfv\'{e}n wings to produce coherent emission,
  or radiate its energy away through processes such as curvature emissions by bunches \citep{Lyutikov2021,Zhang2020A,Wang2022,Mahlmann2022}.
The condition of r$<$r$_{\rm c}$ for the binary degenerate star provides
  an environment of accretion flow or magnetar wind, in which there are enough charged particles for loading in the reconnection,
  giving rise to detectable radio bursts.
Magnetic reconnection also occurs between the magnetar and the 
  close-by planet if the planet is magnetized.
Such reconnection can be modulated by the orbit of the planet, i.e.,
  maximized when the strong magnetic field in the asymmetric planet magnetosphere faces the spinning magnetar \citep[cf.][]{Gao2020}, leading to the 4.605 day quasi-period. 
The second alternative is unipolar induction.
The tidal interaction between a rocky planet and its host star
  could cause volcanic activities,
  which will form unipolar induction and provide plasma around the planet
  for radio maser emission \citep{willes2005,laine2012}.
However, the external plasma from accretion flows or magnetar wind 
  in the binary degenerate system usually contains high energy particles,
  which is still crucial for the detectable high fluence radio bursts.
Thus, regardless of the scenario as we proposed above,
  magnetic reconnection or even unipolar induction,
  the geometrical configuration involving the binary degenerate 
  with a neutron star hosting close orbiting planet 
  (Fig.\ \ref{fig:origin}) would be able to accommodate the periodicity of bursts.

  
Consequently, the bursts from FRB 20121102A may be divided into two classes, 
  arisen from different processes.  
Most of the bursts, in particular, most low-energy bursts,  
  and some high-energy bursts,
  arise from the shocks induced by magnetar blast waves interacting with surrounding plasma. 
Simultaneously, 
  a substantial fraction of high-energy bursts 
  are caused by shock or magnetic interaction between the magnetar and the planet closely orbiting around it,
  which produces more-energetic particles compared to the shock in the surrounding plasma \citep{drain1993}, leading to more luminous radio bursts. 
This may explain the bimodal energy distribution in \cite{Li2021} consisting
  of a log-normal function for low-energyy bursts
  and a generalized Lorentz-Cauchy function for high-energy ones.
Despite the above speculative discussions, more burst properties such as the dispersion measure and rotation measure should be further studied to reveal the origins of these two groups of bursts.


\subsection{Possible formation of FRB 20121102A system} 

Another question is 
  whether or not system of such configuration 
  can be formed. 
It is almost certain that 
  such system cannot be formed by 
  encounter between two degenerate stars, 
  with one carrying a planet. 
The encounter would immediately eject the planet, 
 leaving a binary containing merely two degenerate stars. 
While there are studies \textbf{that have} shown that 
  free floating planet can be captured by a star 
  \citep{Goulinski2018},  
  it is possible that a binary 
  containing two neutron stars or one neutron star and one white dwarf, 
  as a binary would give a larger capturing cross section than a single star \citep{Wang2020}. 
However, such capture will result in a hierarchical system 
   with the planet in an outer orbit around two tightly 
   bounded degenerate stars. 
This kind of systems would not give perodicities reconciling with
  what we have obtained for FRB 20121102A, 
  which requires a hierachy 
  in which the neutron star and the planet form a tight pair 
  with another degenerate star in the outer orbit. 
However, to produce such a hierarchy, 
   it is possible 
   if a binary collide with a star carrying a planet. 
As described in \cite{Li2020}, 
  a binary of such hierarchy can be produced 
  by two possible channels. 
For the system of interest in this work, 
  the initial and final configuration 
  would be either \\ 
(i) $({\rm NS}+{\rm p}) \otimes ({\rm WD}+{\rm X}) \ \longrightarrow 
  (({\rm NS}+{\rm p})+{\rm WD}) \otimes {\rm X}$ \ ; \\ 
(ii) $({\rm X}+{\rm p}) \otimes ({\rm WD}+{\rm NS}) \ \longrightarrow 
  (({\rm NS}+{\rm p})+{\rm WD})  \otimes {\rm X}$ \ , \\ 
where NS, WD, p and X 
  represent neutron star, white dwarf, planet 
  and any kind of \textbf{stellar} object. 
The outcome in these two channels 
  is that the star X would be ejected. 
Their main difference is that the second 
  channel would go through an intermediate state,  
  which determines whether 
  the system will end up with a 
  $(({\rm NS}+{\rm p})+{\rm WD})$ hierarchy 
  or a $(({\rm NS}+{\rm WD})+{\rm p})$ hierarchy
  or something else. 
Stellar encounters are not rare in dense stellar environments,  
  and multiple binary encounter 
  are expected also to occur.  
Accessing the relative yields of the channels requires  
  high precision numerical calculations for  
  the multiple stellar interactions 
  covering a large parameter space. 
Depending on the formation chance, such system may not be frequently seen in other FRB repeaters.
Instead, the binary degenerates or NS - planet system has higher chance to present and may account for the repeating bursts of some FRBs.


\section{Conclusions}
\label{sec:conclu}

Utilizing the PBA method, we analyzed the quasi-periodicity of radio bursts from FRB 20121102A, compiled from the literature. 
We identified a period of 157.1$_{-4.8}^{+5.2}$ day and a candidate period of  4.605$_{-0.010}^{+0.003}$ day, respectively.  
The former is consistent with previous findings 
  and can be explained by the presence of a degenerate binary star system.  
The latter, also the focus of this work, 
  is confirmed by two sets of data.    
It is more clearly seen in energetic bursts with
  fluences larger than $1\times 10^{38}$ erg. 
We attribute this to the possible presence of a planet 
  closely orbiting one of the degenerate stars, presumably a neutron star. 
This dual origin of the bursts is supported by the bimodal distribution of the burst energy 
  observed in FRB 20121102A.  
Shock waves and Alfv\'{e}n wing instabilities induced by magnetic reconnection or unipolar induction
  near the planet 
  are viable mechanisms for producing 
  the radio emissions.  

\begin{acknowledgments}
We acknowledge the referee for professional discussions that help us to improve the manuscript.
This work was supported by National Natural Science Foundation of China
  (NSFC$\#$ 11988101, 42150105, 11725313);
  the National SKA Program of China (Grant No. 2022SKA0120101);
  the Fundamental Research Funds for the Central Universities (Sun Yat-sen University, 2021qntd28, 22qntd3101); 
  and the China Manned Space Project (NO. CMS-CSST-2021-B09 and NO. CMS-CSST-2021-B12). 
YG acknowledges the support from CSST Science Center for the
  Guangdong-Hongkong-Macau Greater Bay Area, SYSU, 
  and the CAS Key Laboratory of FAST;
KW acknowledges the support in part by UK STFC through a 
  Consolidate Grant awarded to UCL MSSL. 
This work has made use of the NASA ADS. 
\end{acknowledgments}



\clearpage

\clearpage

\end{document}